\documentclass[acmtog]{acmart}
\AtBeginDocument{%
  }

\setcopyright{rightsretained}
\acmJournal{TOG}
\acmYear{2024} \acmVolume{43} \acmNumber{6} \acmArticle{237} \acmMonth{12}\acmDOI{10.1145/3687991}

\acmSubmissionID{papers 1174s1}

\usepackage{booktabs} 
\usepackage{subcaption}
\usepackage{multirow}
\usepackage{arydshln}

\usepackage[ruled]{algorithm2e} 

\SetAlFnt{\small}
\SetAlCapFnt{\small}
\SetAlCapNameFnt{\small}
\SetAlCapHSkip{0pt}

\usepackage{kotex}
\usepackage{enumitem}
\usepackage{color}
\definecolor{blue}{rgb}{0,0,0.8}
\definecolor{green}{rgb}{0,0.6,0}
\definecolor{red}{rgb}{0.7,0,0}
\def\blue{\color{blue}}

\def\mat#1{\mathbf{#1}}
\def\vec#1{\mathbf{#1}}

\newcommand{\vf} {\vec{f}}

\newcommand{\vp} {\vec{p}}

\newcommand{\vr} {\vec{r}}

\newcommand{\vx} {\vec{x}}

\newcommand \mT {\mat{T}}

\newcommand{\real} {\mathbb{R}}

\newcommand{\midsepremove}{\aboverulesep = 0mm \belowrulesep = 0mm}

\newcommand{\ignore}[1]{}

\begin{document}

\title{ELMO: Enhanced Real-time LiDAR Motion Capture through Upsampling}

\author{Deok-Kyeong Jang}
\authornote{Equal contribution.}
\email{dk.jang1014@gmail.com}
\orcid{0000-0002-7567-4339}
\affiliation{%
  \institution{MOVIN Inc.}
  \city{Seoul}
  \country{South Korea}
}

\author{Dongseok Yang}
\authornotemark[1]
\email{ds.yang@movin3d.com}
\orcid{0000-0002-4696-3465}
\affiliation{%
  \institution{MOVIN Inc.}
  \city{Seoul}
  \country{South Korea}
}

\author{Deok-Yun Jang}
\authornotemark[1]
\email{dy.jang@movin3d.com}
\orcid{0009-0006-1923-2540}
\affiliation{%
  \institution{MOVIN Inc.}
  \city{Seoul}
  \country{South Korea}
}

\author{Byeoli Choi}
\authornotemark[1]
\email{by.choi@movin3d.com}
\orcid{0000-0003-2347-149X}
\affiliation{%
  \institution{MOVIN Inc.}
  \city{Seoul}
  \country{South Korea}
}

\author{Donghoon Shin}
\email{dh.shin@movin3d.com}
\orcid{0000-0003-2447-5716}
\affiliation{%
  \institution{MOVIN Inc.}
  \city{Seoul}
  \country{South Korea}
}

\author{Sung-Hee Lee}
\authornote{Corresponding author}
\email{sunghee.lee@kaist.ac.kr}
\orcid{0000-0001-6604-4709}
\affiliation{%
  \institution{KAIST}
  \city{Daejeon}
  \country{South Korea}
}

\renewcommand{\shortauthors}{D.-K. Jang et al.}

\begin{abstract}
This paper introduces ELMO, a real-time upsampling motion capture framework designed for a single LiDAR sensor. Modeled as a conditional autoregressive transformer-based upsampling motion generator, ELMO achieves 60 fps motion capture from a 20 fps LiDAR point cloud sequence. The key feature of ELMO is the coupling of the self-attention mechanism with thoughtfully designed embedding modules for motion and point clouds, significantly elevating the motion quality. 
To facilitate accurate motion capture, we develop a one-time skeleton calibration model capable of predicting user skeleton offsets from a single-frame point cloud. Additionally, we introduce a novel data augmentation technique utilizing a LiDAR simulator, which enhances global root tracking to improve environmental understanding.
To demonstrate the effectiveness of our method, we compare ELMO with state-of-the-art methods in both image-based and point cloud-based motion capture. We further conduct an ablation study to validate our design principles. 
ELMO's fast inference time makes it well-suited for real-time applications, exemplified in our demo video featuring live streaming and interactive gaming scenarios. 
Furthermore, we contribute a high-quality LiDAR-mocap synchronized dataset comprising 20 different subjects performing a range of motions, which can serve as a valuable resource for future research.
The dataset and evaluation code are available at {\blue \url{https://movin3d.github.io/ELMO_SIGASIA2024/}}
\end{abstract}




\begin{teaserfigure}
  \centering
  \includegraphics[width=7.0in]{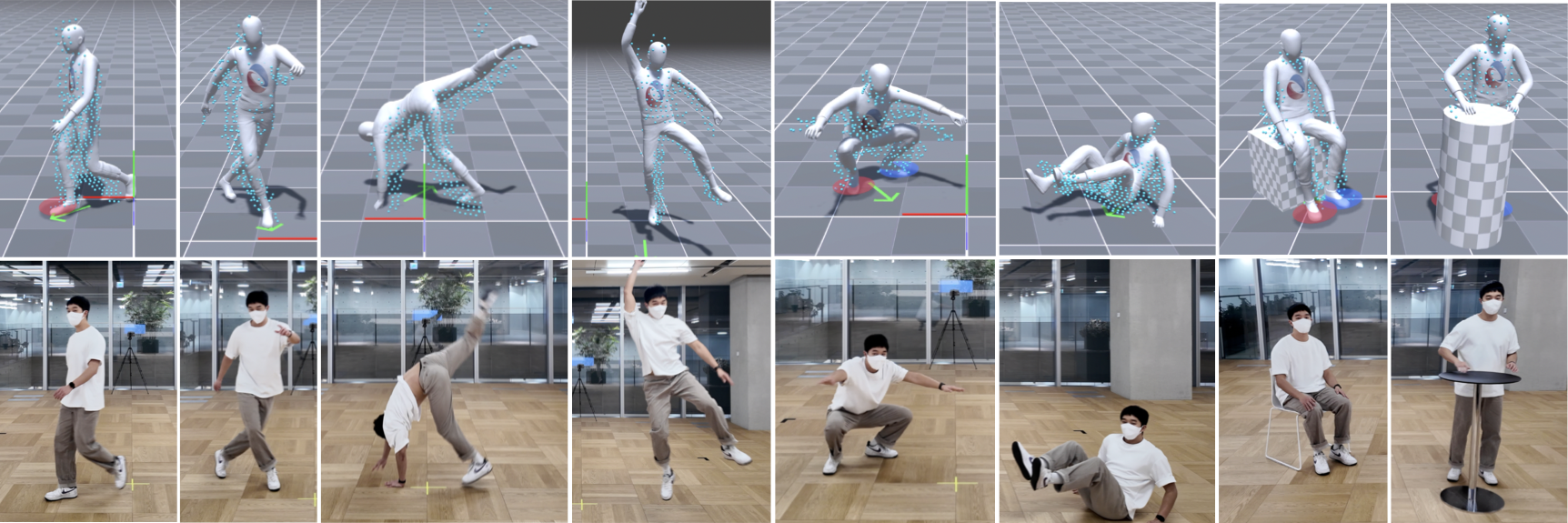}
  \caption{Our ELMO framework enables upsampling motion capture (60 fps) from LiDAR point cloud (20 fps) in real-time.}
 \label{fig:teaser}
\end{teaserfigure}

\maketitle

\section{Introduction}
\label{sec:introduction}
With the increasing need for virtual avatars in 3D content creation and metaverse platforms, real-time motion capture technology is experiencing a notable surge in demand. Despite the strides made by inertial and optical sensors-based solutions with great accuracy, they remain costly and cumbersome for average consumers. Various approaches address this limitation by reducing the number of body-worn sensors \cite{von2017sparse} and exploring more accessible devices such as RGB cameras \cite{wei2010videomocap}. However, RGB cameras and inertial sensors lack explicit information on global translation, leading to drifting in the output. Research utilizing depth sensors, such as RGB-D cameras \cite{zollhofer2014real} and LiDARs \cite{jang2023movin}, demonstrates enhanced global tracking. On the other hand, the measurements from depth cameras suffer from noise and LiDARs' low framerate introduces pose discontinuities in real-time applications, typically requiring higher framerates over 60 frames per second (fps).

To address these limitations, we propose ELMO, a novel real-time upsampling motion capture framework designed to derive 60 fps mocap data from a 20 fps point cloud sequence captured by a single LiDAR sensor. 
The core concept involves using a sequence of sampled point clouds from the past to the current and one future frame to generate three upsampled poses. Our motion generator adopts a conditional autoregressive transformer-based architecture considering past inferred motion and acquired point cloud to establish the relationship between the input point cloud and the output upsampled poses. To overcome self-occlusions in single LiDAR setups, our framework includes a mechanism for sampling a latent vector from a motion prior. This vector is then processed by the generator to predict plausible poses, particularly in scenarios involving occlusions.

To effectively capture the correlation between the LiDAR point cloud and the human body joints, we design motion and point cloud embedding modules such that joint features preserve the skeletal graph structure, a root feature captures global coordinate information, and point features find characteristics of local regions in the point cloud, dubbed \emph{body-patch groups}. Leveraging the self-attention mechanism in our transformer generator, we facilitate the learning of attention between individual body-patch groups and human joints for each embedding feature. This approach significantly enhances the quality of the output motions.

Real-time motion capture relies on accurately tracking global translation, crucial for seamless interaction between avatars and their environment or objects. Acknowledging the challenge of gathering comprehensive data across the capture space, we present a novel data augmentation technique leveraging a LiDAR simulator. We apply global rotations to each original motion clip and fit the SMPL body model~\cite{loper2015smpl} to compute collision points with simulated lasers. Implementing this augmentation technique on our training dataset resulted in a noticeable enhancement in the quality of the mocap results.

Furthermore, we developed a one-time skeleton calibration model that infers user skeleton offsets from a single-frame point cloud acquired while the user is in the A-pose. Skeleton calibration is a fundamental step in motion capture, determining initial joint offsets, global trajectory, and joint rotations.

Figure~\ref{fig:teaser} presents snapshots of real-time mocap results from ELMO. To demonstrate the effectiveness of our framework, we conduct thorough comparisons with state-of-the-art image-based and point cloud-based methods, along with an ablation study to validate our design choices. Additionally, we conduct various experiments, such as testing for global drifting, to verify the essential elements required for accurate motion capture.

To the best of our knowledge, our work is the first real-time upsampling motion capture framework using a single LiDAR. By maintaining low latency, ELMO is well-suited for live application scenarios. Our demo video provides example use cases including live streaming and interactive gaming. The major contributions of our work can be summarized as follows:

\begin{itemize}[leftmargin=*, itemsep=5.0pt]

\item We present the first real-time upsampling motion capture using a single LiDAR, offering low-latency performance for diverse real-time applications.

\item Our novel embedding and generator architecture effectively constructs attention maps between body-patch point groups and human joints, enabling precise upsampling in motion capture. Additionally, we propose a one-time skeleton calibration model to predict user skeleton offsets from a single-frame point cloud.

\item We introduce a new LiDAR-simulation-based data augmentation technique that leverages the unique characteristics of LiDAR sensors to enhance global translation tracking performance. Furthermore, we release a high-quality LiDAR-mocap synchronized dataset featuring 20 subjects performing various actions.

\end{itemize}

\section{Related Work}
\label{sec:related_work}

\subsection{Motion Capture}

Optical and inertial systems stand out in the professional market for their high accuracy. 
However, a shared challenge across these mocap techniques is body-worn sensors that may restrict user motion or shift from their initial positions. Additionally, a time-consuming setup and calibration are required for the quality of captured data. A prominent research focus involves reducing the number of sensors and reconstructing full-body motion from a sparse setup \cite{huang2018deep, jiang2022transformer, winkler2022questsim, lee2023questenvsim, ponton2023sparseposer, yang2024divatrack}. While more accessible than previous methods, they still grapple with inherent issues of wearable sensors.

Consequently, markerless methods \cite{breg1998maps, agu2008multi} have garnered significant attention for their notable advancement in eliminating the need for body-worn sensors. Moreover, they enhance the accessibility of mocap by using widely available devices such as webcams and RGB cameras. Simultaneously, research is underway on both mono \cite{bogo2016smpl, kolotouros2019convolutional, pavlakos2017coarsetofine, wei2022capturing, huang2022neural, bazarevsky2020blazepose, shetty2023pliks, ye2023decoupling, kocabas2020vibe, zhu2022motionbert} and multi-view camera \cite{Amin2013MultiviewPS, Bure2013multiview, Dong2022Multiview} methods. Mono-camera setups offer simplicity while multi-view systems excel in accuracy. While offering supplementary depth information, RGBD solutions \cite{Baak2011depth, VNect_SIGGRAPH2017, Ying2021RgbD} face challenges due to their limited field of view (FoV) and resolutions compared to RGB cameras. This leads to noisy and unstable output poses.

\subsection{LiDAR-based Pose Tracking}

The latest image-based human pose tracking methods \cite{goel2023humans, li2023niki} have shown impressive quality through a two-step process involving 2D keypoint extraction and SMPL body model \cite{loper2015smpl} fitting. However, the accuracy of 3D keypoints and global translation is compromised during pose fitting in 2D image space.

A promising solution is using LiDAR sensors, which provide accurate 3D point cloud data. This method also offers a comprehensive view of the subject's full-body information, not possible with sparse sensor setups. A seminal study by Li et al. \cite{li2022lidarcap} has demonstrated that LiDAR sensors can enhance the quality of captured poses from distances. The following research explores the fusion of LiDAR with IMUs \cite{ren2023lidar} and RGB cameras \cite{cong2023weakly} as a complement for capturing detailed 3D human poses. Recently, Human3D \cite{takmaz20233dseg} has demonstrated remarkable performance in body part segmentation, relying solely on LiDAR point cloud data. Similarly, MOVIN \cite{jang2023movin} performed skeletal motion capture with a single LiDAR sensor. 
Our approach addresses the limitations present in prior works, such as motion jitters and a low frame rate, enhancing the feasibility of LiDAR motion capture frameworks for real-time applications.

\subsection{Neural Generative Models for Human Motion}

Generating natural human motion while minimizing laborious and time-consuming tasks has been among the central focuses in the field of computer graphics. Upon the widespread integration of deep neural networks, researchers developed technologies to generate human motion from various inputs, encompassing low-dimensional control signals, navigation goals, and text prompts. 

Within the realm of neural networks, generative models such as GANs \cite{goodfellow14gan} and VAEs \cite{kingma2013auto} have demonstrated notable success in producing high-quality, natural motion. GANs find application in diverse areas, including character control \cite{wang19gans}, speech-driven gesture generation \cite{ferstl19gesture}, and motion generation from a single clip \cite{li22ganimator}. Conditional VAEs \cite{sohn2015cvae} are employed in motion generation methods that utilize additional constraints such as past motion sequences \cite{ling20character}, motion categories \cite{petrovich21cvae}, and speech \cite{lee19neurips, li20vae}. Recent works have focused on creating a motion embedding space \cite{lee2023same} and leveraging the latent space for tasks like motion in-betweening \cite{tang2023rsmt}, retargeting \cite{li2023ace}, and style transfer \cite{jang2023mocha}. An alternative approach involves using normalizing flow \cite{henter20moglow, aliakbarian2022flag}, enabling exact maximum likelihood. In parallel, building on recent advancements in diffusion models, researchers have extended their application to language \cite{tevet22mdm, zhang22motiondiffuse} and music \cite{tseng22edge}-driven motion synthesis. Methods utilizing deep reinforcement learning frameworks also incorporate VAEs to establish prior distributions for character control \cite{won2022physics}, skills \cite{dou2023c}, unstructured motion clips \cite{zhu2023neural}, and muscle control \cite{feng2023musclevae}.
\section{ELMO Framework}
\label{sec:framework}
\begin{figure*}[ht]
  \centering
  \includegraphics[width=.95\textwidth]{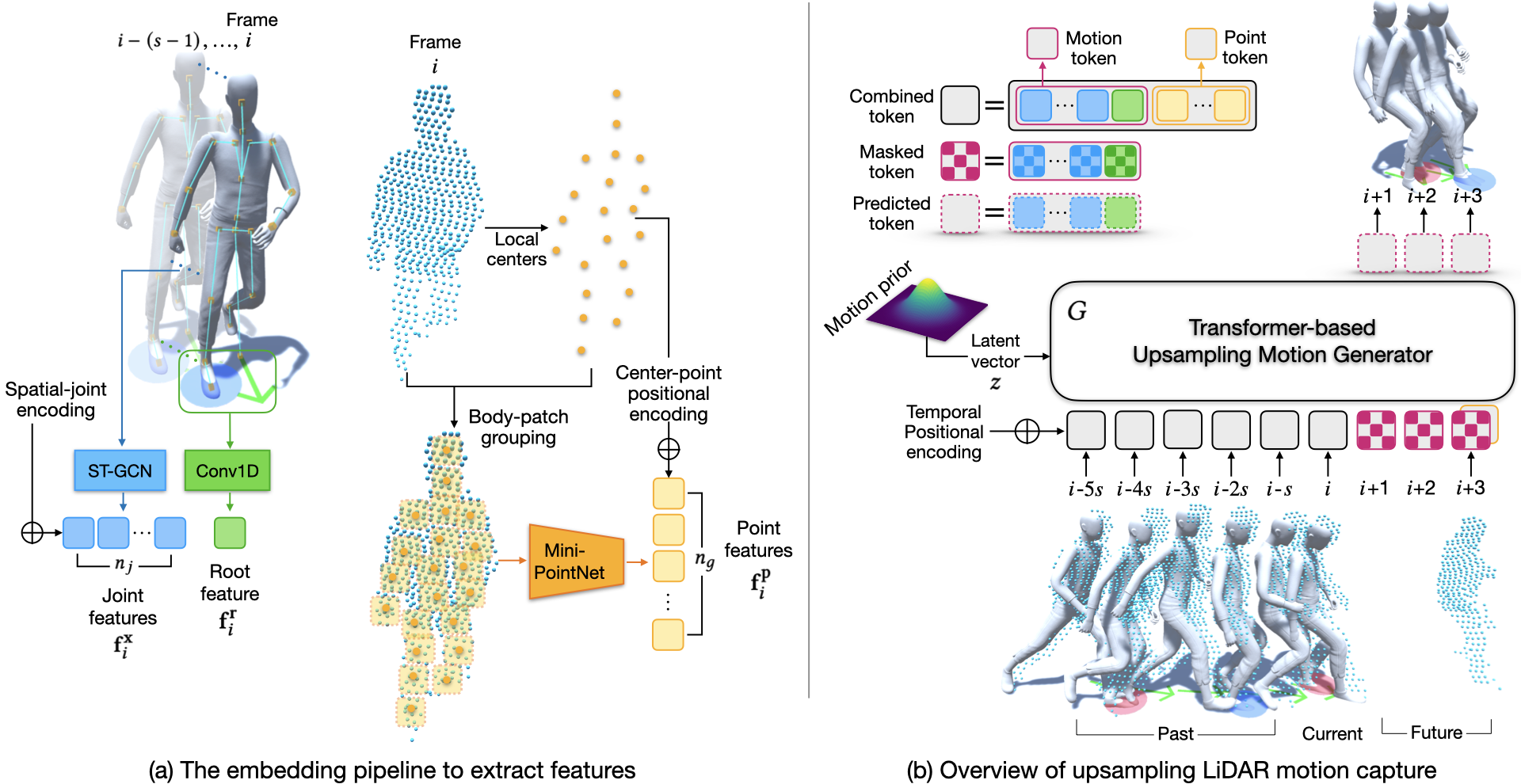}
  \caption{Overall network architectures. (a) Detail of the feature extraction pipeline. (b) Overview of generator for real-time upsampling LiDAR motion capture in run-time.
  }
  \label{fig:overview}
\end{figure*}

Figure ~\ref{fig:overview} (b) illustrates our upsampling motion capture framework from a single-front LiDAR sensor at runtime.
Our framework effectively transforms 20fps LiDAR point cloud input into 60fps motion in real-time, with a latency of one 20fps frame\footnote{Since the input LiDAR operates at 20 fps, when considering a 1 future frame latency, the timestamp will be $i+3$ from the perspective of a 60 fps system.}.
From the inference point of view, the input to our framework is a previously inferred sequence of 60 frames (1 second) of motion, from past frame $i-(6s-1)$ to current frame $i$ (where $s=10$), as well as a sequence of sampled point clouds from the past to the current at timestamps $i-5s$, $i-4s$, $i-3s$, $i-2s$, $i-s$, $i$, and a newly captured point cloud at a future frame $i+3$.
For the output, it generates the next three upsampled poses for frames $i+1$, $i+2$, and $i+3$ at a rate of 60fps. 

Our model is based on an autoregressive conditional transformer-based architecture. At each time frame, we initially extract the features of joints, the root, and points. During the training phase, given the motion sequence as input, the motion distribution encoder $E$ generates a latent vector $z$, which is trained to shape the latent variable $z$ into a Gaussian distribution (Figure~\ref{fig:motion_prior}). 
At the same time, the motion generator $G$ takes the past and current motion sequence, a sampled point cloud sequence of past to one future frame, and the latent vector $z$ as inputs, and generates upsampled poses of the three future frames. 
During inference, the encoder $E$ is discarded. Instead, we pass a randomly sampled latent vector $z$ through the motion generator at each time frame, enabling the generation of plausible poses.

\subsection{Data Representation}
\label{subsec:data_representation}
The input and output of our framework consist of the 3D point cloud and poses.
We start by defining a pose vector for a single frame $i$, which contains a joint vector and root vector. 
Following \cite{jang2023movin}, which suggests the importance of accurate global (world) coordinate information of the character's root for high-quality motion capture, we handle the joint vector and root vector distinctly.
The joint vector $\vx_i$ includes all joint local information including joint local positions, rotations, velocity, and angular velocity with respect to the parent joint, denoted as $\vx_i = [x^t, x^r, \dot{x}^t, \dot{x}^r] \in \mathbb{R}^{n_j \times 15}$, where $x^t \in \mathbb{R}^{n_j \times 3}$, $x^r \in \mathbb{R}^{n_j \times 6}$, $\dot{x}^t \in \mathbb{R}^{n_j \times 3}$, and $\dot{x}^r \in \mathbb{R}^{n_j \times 3}$. Here, $n_j$ denotes the number of joints. The root vector $\vr_i$ represents the global coordinates of each character, including the character's global root position, rotation, velocity, and angular velocity. Additional vectors, such as foot contact, are also incorporated. 
Specifically, we define the root vector as $\vr_i = [r^t, r^r, \dot{r}^t, \dot{r}^r, c] \in \mathbb{R}^{17}$, where $r^t \in \mathbb{R}^3$, $r^r \in \mathbb{R}^6$, $\dot{r}^t \in \mathbb{R}^3$, $\dot{r}^r \in \mathbb{R}^3$, and $c \in \mathbb{R}^2$ for foot contact label. Lastly, the input point cloud vector, representing the global position of human body points, is denoted as $\vp_i \in \mathbb{R}^{n_p \times 3}$, where $n_p$ is the number of points.

Note that the captured motion is 60 fps and the LiDAR input is 20 fps, to prevent confusion, all frames described from now on are in 60 fps.

\subsection{Motion and Point Cloud Embedding}
\label{subsec:embedding}
The embedding process consists of two components: motion embedding and point embedding. The former extracts the spatial and temporal features of the input motion sequence, while the latter groups each point cloud into several local body-patch features.

\paragraph*{\textbf{Motion embedding}}
For motion, we utilize distinct embeddings for joint features and root features, as shown in the left part of Figure ~\ref{fig:overview} (a).
For embedding inputs at frame $i$, we consider frames from $i-(s-1)$ to $i$, denoted as $[\vx_i]_{i-(s-1)}^i = [\vx_{i-(s-1)}, \ldots, \vx_i] \in \mathbb{R}^{s \times n_j \times 15}$ for joint vectors and $[\vr_i]_{i-(s-1)}^i = [\vr_{i-(s-1)}, \ldots, \vr_i] \in \mathbb{R}^{s \times 17}$ for root vectors. Note that we handle motions of length $s$ because our approach involves utilizing point cloud sequences sampled at the interval of $s$ as inputs.

For the joint feature embedding, we use spatial-temporal graph convolutional blocks (STGCN)~\cite{yan2018spatial} to maintain the local joint graph information of the human skeleton as much as possible. The STGCN blocks embed an input sequential joint vectors $[\vx_i]_{i-(s-1)}^{i}$ into a joint features $\vf_i^\vx$ after temporal average pooling as follows:
\begin{equation} \label{eq:joint_embed}
    \vf_i^\vx = \text{TempAvgPool}(\text{ST-GCN}([\vx_i]_{i-(s-1)}^{i})) \in \real^{n_j \times C},
\end{equation}
where $C$ is a feature dimension.

For the root feature, a 1D temporal convolution block is utilized to embed the sequential root vectors $[\vr_i]_{i-(s-1)}^{i}$ to get root feature $\vf_i^\vr$ with equal feature dimension $C$:
\begin{equation} \label{eq:root_embed}
\vf_i^{\vr} = \text{TempAvgPool}(\text{Conv1D}([\vr_i]_{i-(s-1)}^{i})) \in \mathbb{R}^{C}.
\end{equation}

After the motion embedding process, we get a motion feature $[\vf_i^\vx, \vf_i^{\vr}]$ at frame $i$, which achieves temporal alignment with the point cloud data.

\paragraph*{\textbf{Point cloud embedding by body-patch}}
The right part of Figure~\ref{fig:overview} (a) illustrates the points body-patch embedding strategy. Inspired by PointBERT~\cite{yu2022pointbert}, we group each point cloud into several body part patches for each frame $i$. Given an input point cloud of the human body  $\vp_i \in \mathbb{R}^{n_p \times 3}$,  we employ farthest point sampling (FPS) to select $n_g$ central points from the point cloud. By grouping the k-nearest neighbors (k-NN) around these central points, we create $n_g$ distinct local point patches, essentially forming a body-patch cloud with $k$ elements each.
To make these local patches unbiased, we subtract their center coordinates. This step effectively disentangles the structural patterns from the spatial coordinates within the local patches. We then utilize Mini-PointNet~\cite{qi2017pointnet} to project these sub-body-patch clouds into point embeddings.

The Mini-PointNet involves the following steps: Initially, each point within a patch is mapped to a feature vector via a shared multilayer perceptron (MLP). Subsequently, max-pooled features are concatenated to each feature vector. These vectors are then processed through a second shared MLP and a final max-pooling layer, resulting in the body-patch embedding. The overall point cloud embedding process to extract point features $\vf_i^\vp$ is formalized as follows:
\begin{equation} \label{eq:point_embed}
    \vf_i^\vp = \text{Mini-PointNet}(\text{body-patch grouping}(\vp_i)) \in \mathbb{R}^{n_g \times C}
\end{equation}

\subsection{Upsampling Motion Generator}
\label{subsec:generator}
Our motion generator utilizes a conditional autoregressive model, processing past inferred motion and acquired point cloud data as inputs. This establishes a relationship between the current point cloud input and upsampled output poses. By leveraging self-attention within this transformer-based architecture, our approach effectively learns the attention between body-patch point group features and motion features.

\paragraph*{\textbf{Tokenization.}}
We implement a tokenization process for input into a Transformer-based Motion Upsampling Generator, utilizing three distinct token types as illustrated in Figure~\ref{fig:overview}: the Combined Token $\mT_i^{comb}$, Masked Token $\mT_i^{mask}$, and Predicted Token $\mT_i^{pred}$.

The Combined Token $\mT_i^{comb}$ integrates the Motion Token ($\mT_i^{mot}$) and Point Token ($\mT_i^{point}$). The Motion Token consists of joint and root features, added with learnable spatial joint encodings. Conversely, the Point Token, representing the point features of $n_g$ body patches, lacks positional information. Therefore, a two-layer MLP is used to assign positional encodings to each center point of the body-patch groups, which are then added to the body-patch point features:
\begin{equation}
\begin{aligned}
    \mT_i^{mot} &= [\vf_i^\vx, \vf_i^\vr] + \mathcal{P}^\text{spat}, \quad
    \mT_i^{point} = \vf_i^\vp + \mathcal{P}^\text{cent} \\
    \mT_i^{comb} &= [\mT_i^{mot}, \mT_i^{point}],
\end{aligned}
\end{equation}
where $\mathcal{P}^\text{spat}$ denotes spatial joint encoding with learnable parameters, and $\mathcal{P}^\text{cent}$ represents center point positional encoding for body-patch groups.

Next, the Masked Token $\mT_i^{mask}$ is a learnable masking token added with the same spatial positional encoding $\mathcal{P}^\text{spat}$ that, upon passing through the upsampling motion generator, becomes the Predicted Token $\mT_i^{pred}$. The Predicted Token, or pose feature, is employed in reconstructing the upsampled poses. Notably, the Predicted Token (pose feature) corresponding to the masked token frame $i$ represents a pose feature for a single frame, as opposed to the motion feature that deals with the temporal sequence of $i-(s-1)$ to $i$ as used in the combined token. The pose feature also comprises joint features and the root feature.

\paragraph*{\textbf{Transformer-based generator.}}
After the tokenization process, as illustrated in Figure~\ref{fig:overview} (b), we create an input sequence by concatenating the combined tokens $\mT_i^{comb}$ corresponding to frames $i-5s$, $i-4s$, $i-3s$, $i-2s$, $i-s$, and $i$ along with the Point token $\mT_{i+3}^{point}$ for frame $i+3$. Lastly, we pad the masked tokens corresponding to future frames $i+1$, $i+2$ and $i+3$. The temporal positional encoding $\mathcal{P}^\text{temp}$ derived from sinusoidal functions for each time frame is then added to the input sequence.

Given the sampled latent vector $z$, the Upsampling Motion Generator $G$ is an autoregressive model that generates future $i+1$, $i+2$, and $i+3$ pose features (predicted token) at the target frame rate (60fps), conditioned on the input sequence. These 3 pose features are further expanded using expanding modules to obtain the joint vectors $[\tilde{\vx}_{i+1}, \tilde{\vx}_{i+2}, \tilde{\vx}_{i+3}]$ and root vectors $[\tilde{\vr}_{i+1}, \tilde{\vr}_{i+2}, \tilde{\vr}_{i+3}]$. 
We adopt the standard vision transformer for the generator $G$, consisting of multi-headed self-attention layers and FFN blocks. The expanding modules utilize inverse forms of the motion embedding modules. However, a difference is that the expansion is performed separately for the pose of each frame. The overall upsampling generator process is formalized as follows:
\begin{equation} \label{eq:gen}
\begin{aligned}
[\mT_{i+1}^{pred}, \mT_{i+2}^{pred}, \mT_{i+3}^{pred}] =& \\
G([\mT^{comb}, \mT_{i+3}^{point}, &\mT_{i+1}^{mask}, \mT_{i+2}^{mask}, \mT_{i+3}^{mask}] + \mathcal{P}^\text{temp}, z), \\
[\tilde{\vx}_{i+1}, \tilde{\vx}_{i+2}, \tilde{\vx}_{i+3}] = \,\text{De}&\text{-GCN}(\left[\mT_i^{pred}[:n_j]\right]_{i+1}^{i+3}), \\ [\tilde{\vr}_{i+1}, \tilde{\vr}_{i+2}, \tilde{\vr}_{i+3}] = \,\text{De}&\text{-Conv1D}(\left[\mT_i^{pred}[n_j:n_j+1]\right]_{i+1}^{i+3}),
\end{aligned}
\end{equation}
where $\mT^{comb} = [\mT_{i-5s}^{comb}, \mT_{i-4s}^{comb}, \mT_{i-3s}^{comb}, \mT_{i-2s}^{comb}, \mT_{i-s}^{comb}, \mT_{i}^{comb}]$.

\subsection{Motion Prior}
\label{subsec:prior}
\begin{figure}[t]
  \centering
  \includegraphics[width=.80\linewidth]{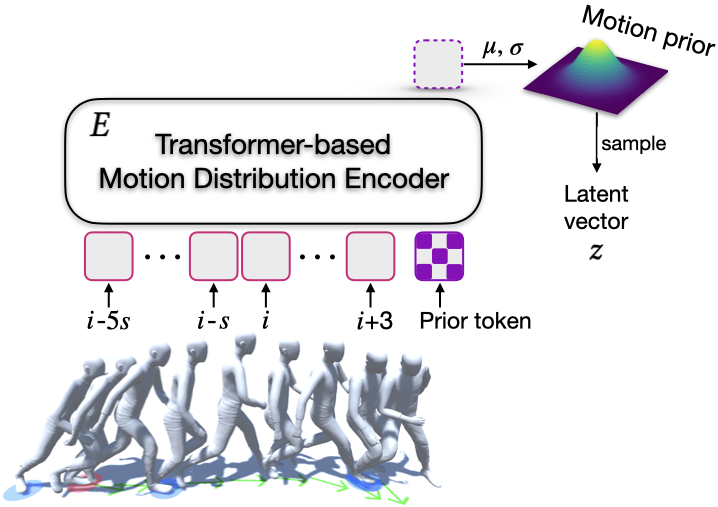}
  \caption{Constructing the motion prior in the training phase.}
  \label{fig:motion_prior}
\end{figure}

The motion prior used to sample the latent vector $z$ is constructed via the motion distribution encoder $E$ as depicted in Figure~\ref{fig:motion_prior}.
Due to self-occlusions among body parts commonly encountered by a single LiDAR sensor, the latent vector $z$, derived from a motion prior, assists the generator in accurately predicting plausible poses.

To effectively capture the spatial-temporal dependencies between past and current poses, we use a transformer architecture, akin to the generator $G$. The motion distribution encoder $E$ processes a learnable prior token $\mT^{prior}$ along with concatenated motion tokens $[\mT_{i-5s}^{mot}, \ldots, \mT_{i+3}^{mot}]$ as its inputs.  These inputs facilitate encoding the parameters of a Gaussian distribution $\mathcal{N}(\mu, \sigma)$. The reparameterization trick is then applied to transform these parameters and obtain the latent vector $z \in \mathbb{R}^{C}$:
\begin{equation} \label{eq:trans_enc}
E(z|[\mT_{i-5s}^{mot}, \ldots, \mT_{i+3}^{mot}] + \mathcal{P}^\text{temp}, \mT^{prior}) = \mathcal{N}(z; \mu, \sigma)
\end{equation}

\subsection{Point Cloud Processing}
\label{subsec:pcdprocessing}
When acquiring point clouds from LiDAR in each frame, extraneous noise points that do not correspond to the human body are often captured due to the surrounding environment. Thus, as a preliminary step, we employ a \textit{Statistical Outlier Removal} algorithm to filter out these irrelevant points. This method uses the mean distance of each point to its neighbors. Points that have a distance significantly larger than the average are considered outliers.

After filtering, our framework is designed to handle an input point cloud with $n_p=384$ points. In cases where the number of points acquired from LiDAR exceeds 384, we apply farthest point sampling (FPS) to reduce the count. Conversely, if the number of points is fewer than 384, we randomly select points and add small noise to generate synthetic points, effectively padding the dataset to the required size.
\section{Training the ELMO Framework}
\label{sec:training}
We train the entire framework end-to-end, optimizing for the loss terms detailed in Section \ref{subsec:loss}. Our training process leverages all available data, both captured and augmented using the techniques outlined in Section \ref{subsec:augmentation}.

\subsection{Data Augmentation}
\label{subsec:augmentation}
\begin{figure}[t]
  \centering  \includegraphics[width=.85\linewidth]{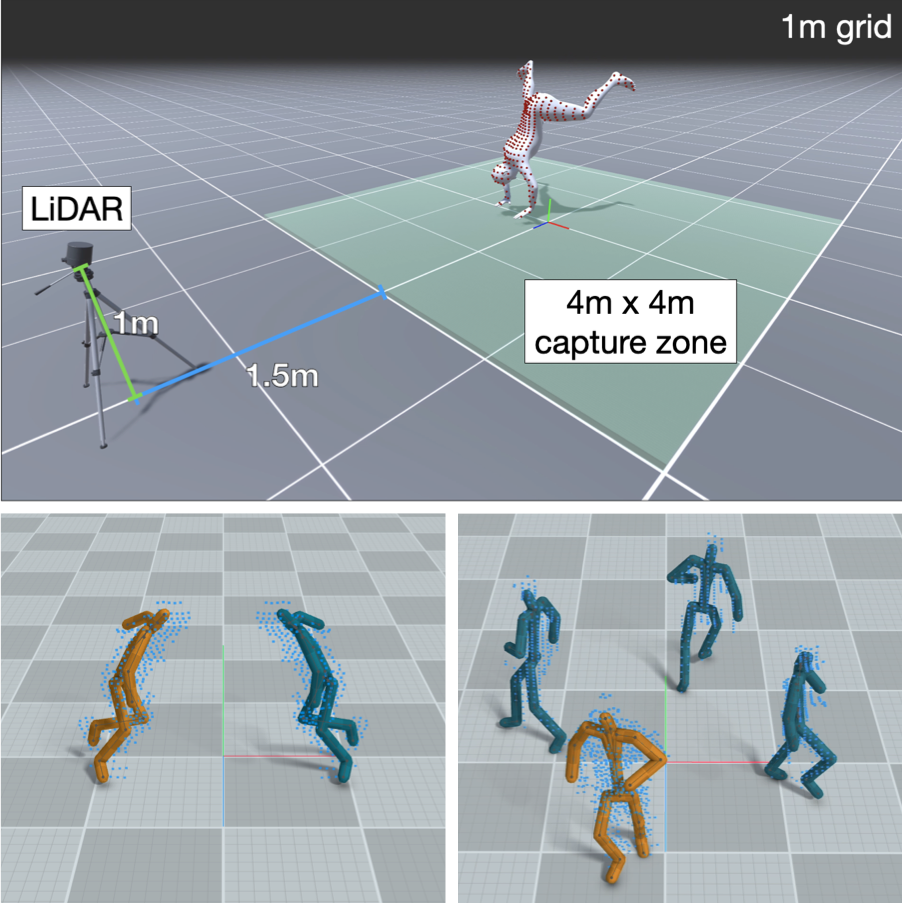}
  \caption{Top: Snapshot of our LiDAR simulator. Red dots represent collision points between the simulated lasers and the body mesh animated with the augmented motion clips. Bottom: Augmentation results using mirroring and simulation for 90°, 180°, and 270° global rotations. The yellow character represents the original data, while the blue characters represent the augmented data.}
  \label{fig:simulation}
\end{figure}

The primary goal of the augmentation is to make the training dataset cover the entire capture space (4$\times$4 meters for our dataset acquisition and experiments), as we use the global coordinate for the root.

We employ two augmentation strategies: mirroring and rotating, as described at the bottom of Figure \ref{fig:simulation}. For mirroring, we double the number of original data by flipping the skeleton and point cloud data. Furthermore, we augment each motion clip by applying global rotations of 90°, 180°, and 270°. However, unlike mirroring, rotating the point cloud around the global origin poses a challenge as a fixed LiDAR would capture different sides of the subject for rotated motion clips. To address this issue, we use a point cloud simulator.

The simulator is implemented with the Unity3D engine and the virtual LiDAR follows the Hesai QT128 specifications \cite{hesaiqt128-link}. To compute collision points with simulated lasers, we use the SMPL body mesh, with its shape parameters manually adjusted to match the subject's skeleton. The top image of Figure~\ref{fig:simulation} shows a snapshot of the resulting simulated point clouds using our simulator. Following our LiDAR placement guidelines to cover a 4m x 4m x 2.5m capture volume, the virtual LiDAR is positioned 3.5 meters from the center of the capture zone (global origin), 1 meter above the ground, and angled 20 degrees downward. During simulation, motion clips run at 60 fps, and point cloud data are captured every 3 frames (20 fps).

\subsection{Losses}
\label{subsec:loss}
The overall framework is trained by minimizing the reconstruction $\mathcal{L}^{rec}$, velocity loss $\mathcal{L}^{vel}$, and KL-divergence $\mathcal{L}^{kl}$ losses. The reconstruction loss comprises joint feature loss on both local and global coordinates and root feature loss. The velocity loss is the difference between consecutive features. The reconstruction and velocity loss are computed for frames $i+1$, $i+2$, and $i+3$. In addition, the KL-divergence loss regularizes the distribution of latent vector $z$ to be near the prior distribution $\mathcal{N}(\boldsymbol{0}, \boldsymbol{I})$.

The total loss function is thus:
\begin{equation} \label{eq:rec}
\begin{aligned}
    \mathcal{L}^{total} = & \mathcal{L}^{rec}|_{i+1}^{i+3} + w_{vel}\mathcal{L}^{vel}|_{i+1}^{i+3} + w_{kl}\mathcal{L}^{kl} \\
    \mathcal{L}^{rec}_i = &\| \tilde{\vx}_i - \vx_i \|_1 + \| FK(\tilde{\vx}_i)- FK(\vx_i) \|_1 + \| \tilde{\vr}_i - \vr_i \|_1 \\
    \mathcal{L}^{vel}_i = &\| V(\tilde{\vx}_i) - V(\vx_i) \|_1 + \| V(FK(\tilde{\vx}_i))- V(FK(\vx_i)) \|_1 \\
    +& \| V(\tilde{\vr}_i) - V(\vr_i) \|_1,
\end{aligned}
\end{equation}
where $V(\vx)= \frac{\vx^0-\vx^1}{h}$, $V(FK(\vx))= \frac{FK(\vx^0)-FK(\vx^1)}{h}$, $h$ is time step, and $w_{kl}$, $w_{delta}$ are relative weights. $FK$ represents forward kinematics.

\subsection{Implementation details}
The AdamW optimizer was used over 30 epochs with a learning rate of $10^{-4}$. The loss weights $w_{vel}$ and $w_{kl}$ were both set to 1. In the embedding module, the ST-GCN and 1D convolution comprise one layer along with temporal pooling to extract joint features and the root feature. We split the input point cloud into 32 body-patch groups, which are input to Mini-PointNet, composed of one set abstraction layer.
The upsampling motion generator $G$ comprises 3 vision transformer layers with 384 channels and 8 heads. The motion distribution encoder $E$ has the same architecture as the generator. The expanding module has an architecture symmetric to the embedding module.
To prevent covariate shifts during autoregressive inference, we set the prediction length to 20 frames for training. Scheduled sampling was also utilized to make the model robust to its own prediction errors, enabling long-term generation. Training took around three days using two 24GB RTX4090 GPUs.
\section{Skeleton Calibration Model}
\label{sec:calibration}

Commercial motion capture systems measure bone lengths directly \cite{xsens-link} or optimize from pre-programmed marker sets \cite{optitrack-link}. Prior data-driven methods continuously predict body shape and motion together from input sequences \cite{jiang2023egoposer, ren2024livehps}. In line with commercial setups, we devise a one-time skeleton calibration model to precede the capture session. The model predicts the user skeleton offsets from a single-frame point cloud, acquired while the user is in the A-pose.

\subsection{Dataset Synthesis}
\begin{figure}[t]
  \centering
  \includegraphics[width=.95\linewidth]{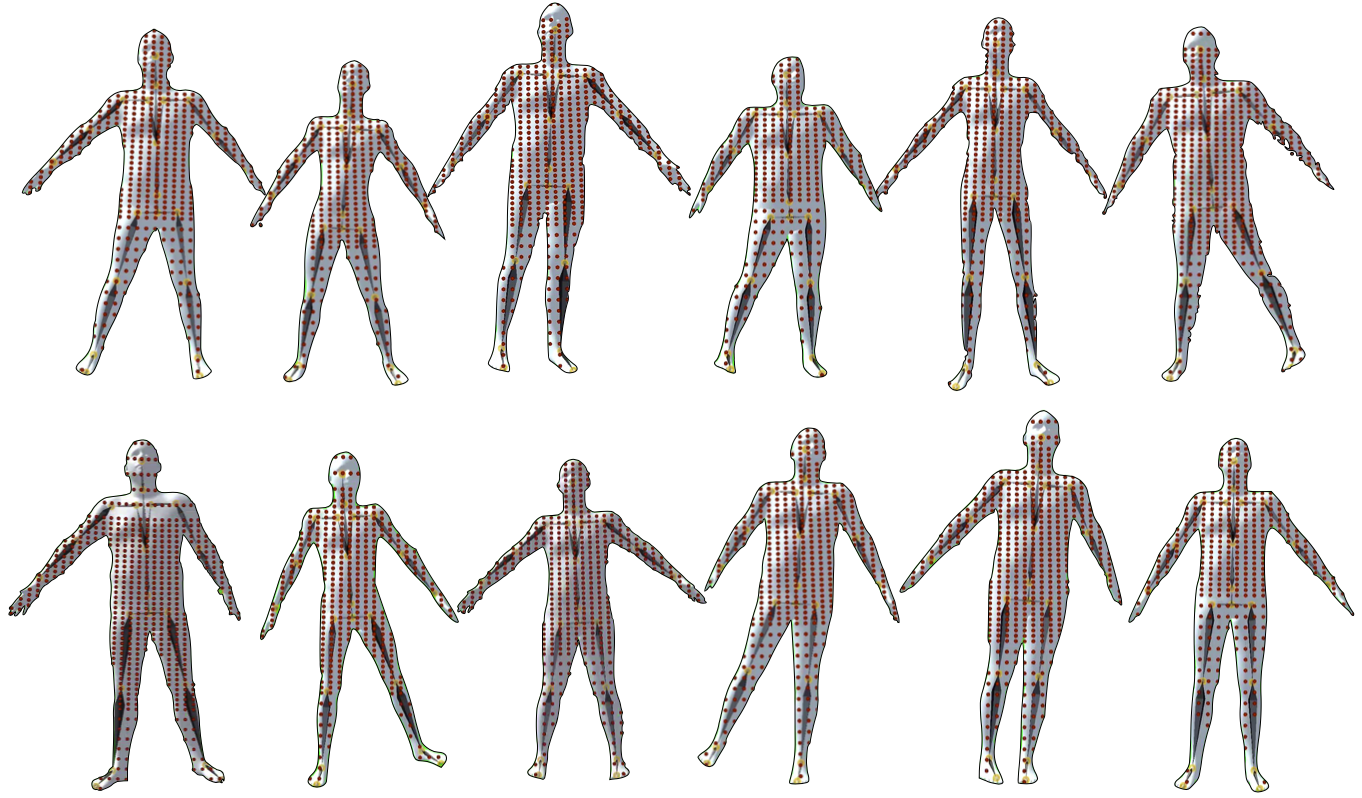}
  \caption{Samples of random SMPL body meshes in A-poses with corresponding skeletons and simulated point clouds.}
  \label{fig:Aposesimulation}
\end{figure}

To accommodate diverse body shapes among users, we generate a synthetic dataset comprising 50,000 pairs of an A-pose point cloud and initial joint offsets using our LiDAR simulator (Sec.\ref{subsec:augmentation}). Each data pair was generated through a randomized process to ensure the model's efficacy in real-world scenarios, despite being trained and validated solely on synthetic data. SMPL shape parameters of each subject are sampled in [-2, 2], covering 95\% of the SMPL shape space. To enhance robustness against deviations in LiDAR placements in practice, we introduce random positional and rotational offsets to the virtual LiDAR, with maximums of 10cm and 5°, respectively. To further improve the model's robustness against variations in user A-poses, we apply random joint rotations: shoulder joints z-axis [65°, 45°], shoulder joints x-axis [-30°, 30°], hip joints y-axis [-30°, 20°], and hip joints z-axis [-20°, 4°]. In the final stage, the obtained SMPL joint offsets are retargeted to our skeleton hierarchy. Figure \ref{fig:Aposesimulation} presents example images of SMPL body mesh with random shape parameters in random A poses, corresponding skeletons, and simulated point clouds. The dataset was split for training and validation in an 80:20 ratio, resulting in 40,000 and 10,000 pairs, respectively.

\subsection{Calibration Model \& Training}
Our skeleton calibration model is a simple 6-layer multi-layer perceptron with bias disabled. It takes as input a flattened vector of 384 3D points sorted by their global height, $x\in\mathbb{R}^{384\times3}$. A preprocessing step, detailed in Sec. \ref{subsec:pcdprocessing}, is applied to sampled A-pose point clouds to ensure a fixed number of input points. Additionally, we apply noise ($X \sim \mathcal{N}(1\text{cm},\,1^{2}$)) to each position channel (x, y, z) of simulated points to enhance the robustness of our model in handling real-world noise induced by factors such as user outfits and hairstyles. Given this input, our model predicts skeleton offsets, including initial hip height and 20 joint offsets, represented as $y\in\mathbb{R}^{1+20\times3}$. Training is conducted for 220 epochs to minimize mean squared error, utilizing the Adam optimizer with a learning rate of $5.0\times10^{-6}$, and employing a batch size of 2048.
\section{Evaluation and Experiments}
\label{sec:Evaluation}

To assess the effectiveness of our framework, we conduct both quantitative and qualitative evaluations, comparing it with state-of-the-art (SOTA) methods in human motion tracking.
Furthermore, we examine how the self-attention mechanism operates between body-patch point groups and joint features by constructing attention maps. 
Lastly, to showcase the real-time capability of ELMO, we present live-streaming scenarios with both single and multi-person setups and an interactive shooting game. 
For visual animation results, please refer to the supplementary video.

\subsection{Datasets}
\label{subsec:dataset}
\midsepremove
\begin{table}[t]
\caption{Motion categories in ELMO dataset.}
\centering
\resizebox{.95\columnwidth}{!}{
\begin{tabular}{|l|}
\midrule
\textbf{Range of Motion} \\
\midrule
Freely rotating an individual joint/joints in place \& while moving \\
\midrule
\textbf{Static Movements} \\
\midrule
T-pose, A-pose, Idle, Look, Roll head \\
Elbows bent up \& down, Stretch arms \\
Bow, Touch toes, Lean, Rotate arms \\
Hands on waist, Twist torso, Hula hoop \\
Lunge, Squat, Jumping Jack, Kick, Lift knee \\
Turn in place, Walk in place, Run in place \\
Sit on the floor \\
\midrule
\textbf{Locomotion} \\
\midrule
Normal walk, Walk with free upper-body motions \\
Normal Jog, Jog with free upper-body motions \\
Normal Run, Run with free upper-body motions \\
Normal Crouch, Crouch with free upper-body motions \\
Transitions with changing pace \\
Moving backward with changing pace \\
Jump (one-legged, both-legged, running, ...) \\
\midrule
\end{tabular}
}
\label{tab:datacomposition}
\end{table}

We construct the ELMO dataset, a high-quality, synchronized single LiDAR-Optical Motion Capture-Video dataset featuring 20 subjects (12 males / 8 females, height$_{cm}\ h \sim \mathcal{N}(170.66,\ 7.90^{2})$,\ $155 \leq h \leq 180$). Our objective is to capture a wide range of motions, styles, and body shapes. We utilize a 4$\times$4 meter space, Hesai QT128 LiDAR, and an Optitrack system equipped with 23 cameras. 
The point cloud and mocap data were recorded at 20 and 60 fps, respectively. We split the 20 subjects into training and test sets with 17 and 3 subjects.

To capture point clouds from diverse distances and angles of a single LiDAR, we subdivided the capture zone into four separate subspaces and defined four viewing directions: forward, backward, left, and right (+z, -z, +x, -x). For each action category, participants executed the same action in 16 combinations of zone and viewing directions. To ensure our model accommodates diverse human poses, we initially captured a Range of Motion (ROM), including individual movements of each joint, along with their free combinations. Subsequently, we recorded in-place motions, followed by diverse locomotions involving different velocities, directions, and styles. Table \ref {tab:datacomposition} presents details on action labels comprising the ELMO dataset. The captured LiDAR point cloud and .bvh mocap files were precisely synchronized by our preprocessing pipeline.

We additionally tested our algorithm on the MOVIN dataset \cite{jang2023movin} for quantitative evaluation. It consists of 10 subjects including 4 males and 6 females performing actions such as Walking, Jogging, Jumping, and Sitting on the floor. For the MOVIN dataset, we use 8 subjects for the training and 2 subjects for the test set.

For qualitative comparison, we also created a wild test dataset consisting of 3 subjects. To compare the performance of markerless motion capture without a suit, we built a synchronized single LiDAR-IMU-based Motion Capture-Video dataset. The IMU-based Motion Capture system used Xsens Awinda~\cite{awinda}.

\subsection{Quantitative Evaluation}
\label{subsec:quantitative}
We quantitatively compare our results with SOTA image-based pose tracking methods including VIBE \cite{kocabas2020vibe}, MotionBERT \cite{zhu2022motionbert}, and NIKI \cite{li2023niki}. 
Additionally, MOVIN \cite{jang2023movin} serves as a primary comparison, utilizing the same LiDAR input as our framework. The quantitative evaluation assesses the methods using both MOVIN and our new ELMO dataset.

Quantitative measures are defined in terms of the mean joint (J) and pelvis (P) position/rotation/linear velocity/angular velocity errors (M*PE, M*RE, M*LVE, M*AVE). We measure joint errors for the fixed pelvis and separately measure pelvis errors, as image-based methods cannot explicitly track global trajectory. The accuracy of pelvis prediction significantly influences overall motion quality, as errors at the root joint propagate along the kinematic chain of the joint hierarchy. For comparison, we retarget the SMPL output of NIKI to our skeleton hierarchy \footnote{Retargeting is performed to enable comparison within the same skeleton topology. However, this process might introduce some unfairness.}.

\begin{table}[t]
\small
\caption{Quantitative comparison between baseline and ELMO models on the test splits of the ELMO and MOVIN datasets. ``w/ dup'' signifies output upsampled via duplication, while ``w/ interp'' denotes via interpolation. MOVIN$^{\dag}$ indicates the model retrained with the ELMO dataset. To ensure a fair comparison on the MOVIN dataset, results from ELMO (Ours) are downsampled to 20 fps.}

\begin{subtable}[h]{0.5\textwidth}
\centering
\begin{tabular}{lcccc}
\midrule
& MJPE${}_{cm}$ & MJRE${}^\circ$ & MJLVE & MJAVE \\
\midrule
NIKI & $14.30$ & $18.04$ & $1.41$ & $2.35$ \\
MOVIN$^{\dag}$ w/ dup & $\underline{7.03}$ & $11.87$ & $1.32$ & $1.65$ \\
MOVIN$^{\dag}$ w/ interp & $7.05$ & $\underline{11.87}$ & $\underline{1.08}$ & $\underline{1.45}$ \\
ELMO (Ours) & $\textbf{4.86}$ & $\textbf{10.41}$ & $\textbf{0.38}$ & $\textbf{0.77}$ \\
\midrule
& MPPE${}_{cm}$ & MPRE${}^\circ$ & MPLVE & MPAVE \\
\midrule
MOVIN$^{\dag}$ w/ dup & $8.88$ & $6.27$ & $1.55$ & $1.31$ \\
MOVIN$^{\dag}$ w/ interp & $\underline{8.73}$ & $\underline{6.56}$ & $\underline{0.67}$ & $\underline{1.54}$ \\
ELMO (Ours) & $\textbf{4.08}$ & $\textbf{5.08}$ & $\textbf{0.20}$ & $\textbf{0.38}$ \\
\midrule
\end{tabular}
\caption{ELMO dataset.}
\label{tab:quansota_elmo}
\end{subtable}

\begin{subtable}[h]{0.5\textwidth}
\centering
\begin{tabular}{lcccc}
\midrule
& MJPE${}_{cm}$ & MJRE${}^\circ$ & MJLVE & MJAVE \\
\midrule
VIBE & $10.86$ & $18.39$ & $2.39$ & $3.16$ \\
MotionBERT & $10.62$ & $18.05$ & $\underline{1.75}$ & $\underline{2.24}$ \\
NIKI & $12.32$ & $17.21$ & $1.90$ & $3.32$ \\
MOVIN & $\underline{6.21}$ & $\textbf{10.12}$ & $1.89$ & $2.75$ \\
ELMO (Ours) & $\textbf{4.77}$ & $\underline{11.00}$ & $\textbf{0.92}$ & $\textbf{1.53}$ \\
\midrule
& MPPE${}_{cm}$ & MPRE${}^\circ$ & MPLVE & MPAVE \\
\midrule
MOVIN & $\underline{4.42}$ & $\underline{11.64}$ & $\underline{2.46}$ & $\underline{4.94}$ \\
ELMO (Ours) & $\textbf{4.42}$ & $\textbf{6.49}$ & $\textbf{0.49}$ & $\textbf{0.91}$ \\
\midrule
\end{tabular}
\caption{MOVIN dataset.}
\label{tab:quansota_movin}
\end{subtable}

\label{tab:quansota}
\end{table}

\begin{table}[t]
\caption{Quantitative measures of ablation models of future frame input and dataset augmentation.}
\small

\begin{subtable}[h]{0.5\textwidth}
\centering
\begin{tabular}{lcccc}
\midrule
& MJPE${}_{cm}$ & MJRE${}^\circ$ & MJLVE & MJAVE \\
\midrule
ELMO$_{20}$ w/ interp & $\underline{5.05}$ & $\underline{11.45}$ & $0.54$ & $0.91$ \\
\hdashline
ELMO baseline & $6.33$ & $12.48$ & $0.48$ & $0.84$ \\
+ 1 future frame & $5.33$ & $11.61$ & $\underline{0.39}$ & $\underline{0.78}$ \\
+ Augmentation (Ours) & $\textbf{4.86}$ & $\textbf{10.41}$ & $\textbf{0.38}$ & $\textbf{0.77}$ \\
\midrule
& MPPE${}_{cm}$ & MPRE${}^\circ$ & MPLVE & MPAVE \\
\midrule
ELMO$_{20}$ w/ interp & $\underline{4.10}$ & $\underline{5.13}$ & $0.19$ & $0.82$ \\
\hdashline
ELMO baseline & $4.97$ & $6.39$ & $0.28$ & $ 0.45$ \\
+ 1 future frame & $5.05$ & $5.20$ & $\textbf{0.19}$ & $\underline{0.39}$ \\
+ Augmentation (Ours) & $\textbf{4.08}$ & $\textbf{5.08}$ & $\underline{0.20}$ & $\textbf{0.38}$ \\
\midrule
\end{tabular}
\caption{ELMO dataset.}
\label{tab:ablation_elmo}
\end{subtable}

\begin{subtable}[h]{0.5\textwidth}
\centering
\begin{tabular}{lcccc}
\midrule
& MJPE${}_{cm}$ & MJRE${}^\circ$ & MJLVE & MJAVE \\
\midrule
ELMO baseline & $6.14$ & $\underline{11.31}$ & $1.02$ & $1.58$ \\
+ 1 future frame & $\underline{5.39}$ & $11.65$ & $0.98$ & $\underline{1.57}$ \\
+ Augmentation (Ours) & $\textbf{4.77}$ & $\textbf{11.00}$ & $\textbf{0.92}$ & $\textbf{1.53}$ \\
\midrule
& MPPE${}_{cm}$ & MPRE${}^\circ$ & MPLVE & MPAVE \\
\midrule
ELMO baseline & $5.03$ & $8.66$ & $0.59$ & $1.01$ \\
+ 1 future frame & $5.08$ & $\underline{6.69}$ & $\underline{0.55}$ & $0.95$ \\
+ Augmentation (Ours) & $\textbf{4.42}$ & $\textbf{6.49}$ & $\textbf{0.49}$ & $\textbf{0.91}$ \\
\midrule
\end{tabular}
\caption{MOVIN dataset.}
\label{tab:ablation_movin}
\end{subtable}

\label{tab:ablation}
\end{table}

\paragraph*{\textbf{SOTA comparison on ELMO dataset.}}
The upper section of Table \ref{tab:quansota} compares our model's quantitative measures with baselines on the ELMO dataset. Given that MOVIN operates at 20 fps, we retrain it with a downsampled ELMO dataset, and its outputs are upsampled to 60 fps using duplication (w/ dup) and interpolation (w/ interp) for comparison.

For both joint and pelvis measures, our ELMO significantly outperformed the baselines. Compared to MOVIN upsampled with interpolation, the best among the baselines, the improvements are particularly notable in position measures, with a decrease of 2.19 cm in MJPE and 4.65 cm in MPPE. Additionally, our model showed performance increases ranging from a minimum of 47\% (MJAVE) to a maximum of 75\% (MPAVE), demonstrating its strength in capturing natural, non-linear pose transitions in both temporal and spatial spaces. This is further illustrated by the example outputs in panel (a) of Figure \ref{fig:quancombined}, where ELMO generates poses with greater accuracy (sitting, upper row) and faster reactions to input point clouds (running, bottom row).

\paragraph*{\textbf{SOTA comparison on the MOVIN dataset.}}
Since the MOVIN dataset only provides 20 fps motion capture (mocap) data, we retrain ELMO on a synthetic 60 fps MOVIN dataset using interpolation, and the outputs are then downsampled to 20 fps for comparison.

The bottom part of Table \ref{tab:quansota} compares quantitative measures of our model with baselines on the MOVIN dataset. Overall, our ELMO outperformed the baselines in joint measures, except for MJRE. Despite a slight increase in MJRE by $0.88^{\circ}$ compared to MOVIN, ELMO showed a notable improvement in MJPE by 1.44cm, indicating superior accuracy in joint positions. We observed that ELMO especially performed better during occlusions, as shown in the bottom row of the panel (b) of Figure \ref{fig:quancombined}, where the MOVIN output incorrectly lifted the opposite arm. Moreover, ELMO significantly outperformed MOVIN in MPRE, surpassing it by $5.15^{\circ}$; the upper row of Figure \ref{fig:quancombined} shows an instance where MOVIN exhibited a global y-axis flip in the output.

\paragraph*{\textbf{Ablation study.}}
To assess the impact of our novel upsampling motion generator (Sec.~\ref{subsec:generator}) and data augmentation (Sec.~\ref{subsec:augmentation}), we conducted an ablation study in (a) ELMO and (b) MOVIN datasets. 
The baseline ELMO corresponds to the basic setup of our model , which predicts the poses of 3 future frames from the point cloud input of the current frame.
In addition, we include a 20 fps output model without the upsampling scheme. For comparison within the ELMO dataset, the 20 fps outputs are upsampled using interpolation, denoted as ELMO$_{20}$ w/ interp.

Table \ref{tab:ablation} presents metrics from ablation models. Testing on the ELMO dataset revealed that incorporating a future frame input generally decreased errors, except for MPPE, which maintained comparable performance with a 0.08cm gap. This enhancement notably improved the model's accuracy in local body pose transitions compared to the baseline. Dataset augmentation yielded modest improvements in velocity errors but notably enhanced position and rotation errors for both body joints and pelvis.
In the case of ELMO$_{20}$ w/ interp., since the output poses are generated to match the point cloud input fps, the position and rotation errors are relatively low, but the overall velocity error is high.

\begin{figure}[t]
\centering
\includegraphics[width=0.8\linewidth]{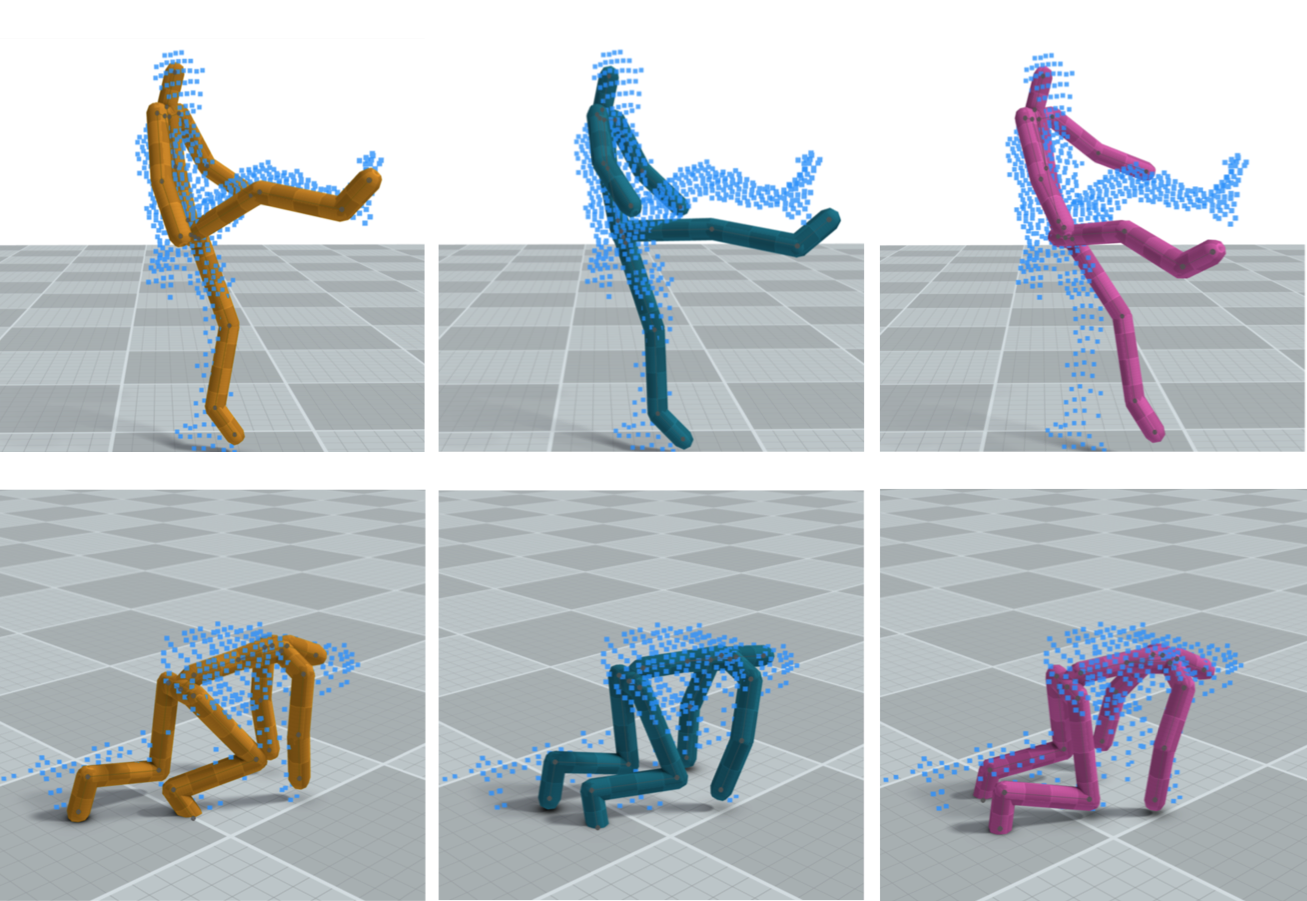}
\caption{Samples of offline outputs of ablation models. From left to right: ELMO with future frame input and data augmentation (Yellow), only with future frame input (Blue), and the baseline (Pink).}
\label{fig:ablation}
\end{figure}

Figure \ref{fig:ablation} displays examples from the ablation models. In the first row, adding a future frame enhanced leg pose accuracy during a kicking motion, and augmentation further improved performance, aligning the output with the point cloud input. In the bottom row, data augmentation increased robustness during occlusions, enhancing leg details while crawling.

\paragraph*{\textbf{Inference time comparison.}}
For real-time applications, achieving a minimum of 60 fps is crucial. 
Table \ref{tab:totaltime} outlines the elapsed times for methods on a Windows Laptop PC (i9-13900HX, RTX4080: TGP 175W).
NIKI's inference time of 14 ms falls within the acceptable limit for 60 fps. However, the total time for bounding box detection, a prerequisite step before pose generation, is 454 ms, rendering it unsuitable for real-time applications. 
MOVIN demonstrates an inference time of 48 ms. When combined with the point cloud capture and processing time, the total time increases to 54 ms, barely meeting the 20 fps framerate. 
In contrast, although ELMO generates poses from the previous two frames to the current frame at a 60Hz rate, resulting in an input latency of 33ms, the inference time is only 5ms, bringing the total end-to-end elapsed time to just 44ms.

The reason our inference time is significantly faster compared to MOVIN lies in the overall structure of our model architecture. Specifically, in the point embedding section, MOVIN simply utilizes PointNet++, whereas our method employs body-patch grouping and Mini-PointNet to efficiently extract point features for each group.

\begin{table}[h]
\caption{Total elapsed times for different frameworks.}
\centering
\small
\begin{tabular}{lc}
\midrule
Framework  & Time$_{ms}$ \\
\midrule
NIKI inference & 14 \\
+ bounding box detection & 454 \\
\midrule
MOVIN inference & 48 \\
+ point cloud capture \& process & 54 \\
\midrule
ELMO inference & 5 \\
+ latency & 38 \\
+ point cloud capture \& process & 44 \\
\midrule
\end{tabular}
\label{tab:totaltime}
\end{table}

\paragraph*{\textbf{Skeleton offset error.}}
Table \ref{quan:calib} displays the average bone length and direction errors of our skeleton calibration model (Sec.~\ref{sec:calibration}) evaluated on a test set comprising 7 subjects with heights of 155, 160, 168, 171, 177, and 179cm sampled from ELMO dataset. Our model achieves low errors, with an average joint length error of 1.52cm and direction error of 0.22°. Figure \ref{fig:subj_cali} and supplementary video demonstrate the calibration model's robust performance across users with diverse body shapes, outfits, and hairstyles. The model accurately predicts initial offsets, facilitating precise fitting of the scaled mesh to the captured point cloud.

\begin{table}[h]
\caption{Average offset length and direction errors by body parts (number of joints in brackets).}
\small
\begin{tabular}{lccccc}
\midrule
& Hips (1) & Spine (4) & Arm(8) & Leg (8) & Total (21) \\
\midrule
length error$_{cm}$ & $2.41$ & $1.22$ & $1.68$ & $1.39$ & $1.52$ \\
direction error° & $-$ & $0.11$ & $0.34$ & $0.17$ & $0.22$ \\
\midrule
\end{tabular}
\label{quan:calib}
\end{table}

\subsection{Qualitative Evaluation}
\label{subsec:qualitative}
We qualitatively compare our real-time output of ELMO with Xsens Awinda~\cite{awinda}, MOVIN, and NIKI~\footnote{The outputs from NIKI lack root transformation, so we copy the root transformation from the Xsens results to the NIKI output.} for the wild test dataset (Sec.~\ref{subsec:dataset}). Please refer to the supplementary video for a visual comparison.

\paragraph*{\textbf{Comparison with Xsens and SOTA methods.}}
Figure~\ref{fig:qualitative} presents the ground truth sequence and real-time output sequence of ELMO, Xsens Awinda, MOVIN, and NIKI. Overall, NIKI performs poorly across all scenarios, containing severe jitters and inaccurate poses. 
Xsens shows weakness in height change and global drifting, exemplified in jumping (4$^{th}$ row) and rapid turning (6$^{th}$ row). Moreover, the inherent limitation of acceleration-based tracking results in inaccurate overall global positions. 

MOVIN exhibits significant jitter in the output poses and frequently struggles to match the input point cloud (1$^{st}$, 2$^{nd}$, and 5$^{th}$ rows). As the method runs at 20 fps, the output poses often lose continuity, especially for relatively fast and continuous motions, including rotating the foot (5$^{th}$ row) and high kick (1$^{st}$ row). In addition, the method sometimes suffers a y-axis flip for global orientation as in ballet turn ({3$^{rd}$ row), making the character face the wrong direction. Furthermore, since the output is at 20 fps, severe frame drop can be observed when visualized in a 60 Hz environment.

In contrast, ELMO achieves natural and continuous 60 fps motions akin to Xsens while accurately capturing the global root trajectory. Furthermore, ELMO excels in capturing precise details such as body part contacts and producing plausible motions even in the presence of occlusion.

\begin{figure}[t]
\centering
\includegraphics[width=.8\linewidth]{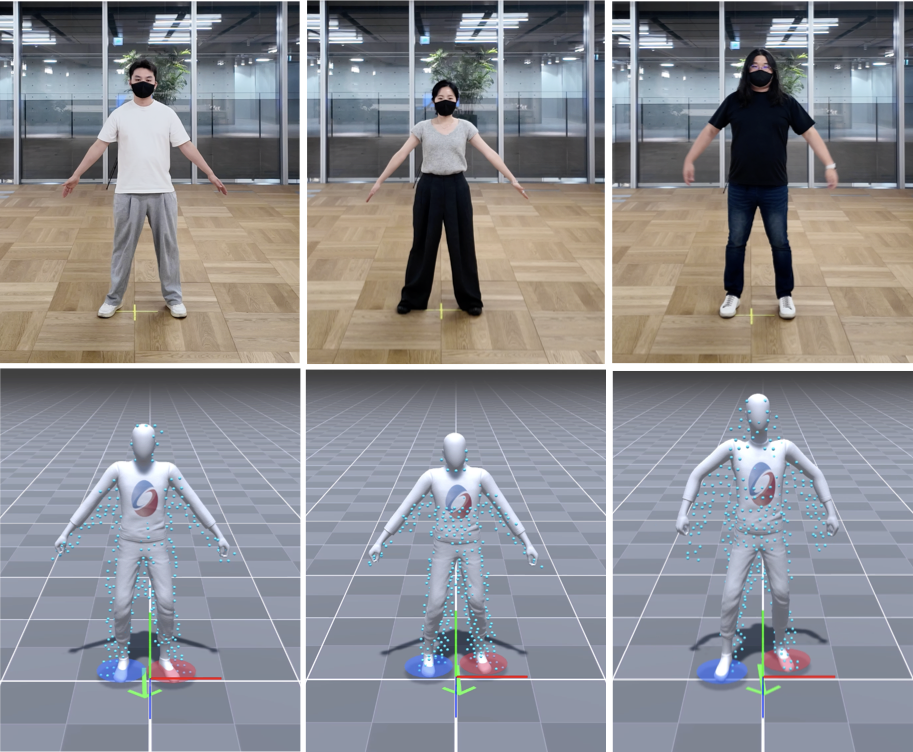}
\caption{Skeleton calibration results for wild input. From left to right: 171 cm male, 157 cm female, and 182 cm male.}
\label{fig:subj_cali}
\end{figure}

\paragraph*{\textbf{Skeleton Calibration.}}
For the qualitative evaluation of the skeleton calibration model, we conducted a wild test with three subjects: a 157 cm female, a 171 cm male, and an 182 cm male. After acquiring a single-frame point cloud while each subject was in the A-pose at origin, our model predicted the user skeleton offsets. Figure~\ref{fig:subj_cali} demonstrates that our model successfully predicted the skeleton offsets for each subject.

\paragraph*{\textbf{Global drifting.}}
Global drifting in motion capture leads to inaccuracies over time, significantly affecting the precision and reliability of motion data. In long sequences, these errors accumulate, causing substantial positional deviations. 
To compare the global drifting issues, we conducted an experiment where the subjects initiated dynamic motion from the origin, continued for 1-2 minutes, and then returned to the origin. The objective was to determine whether they precisely returned to the same position.

As depicted in Figure~\ref{fig:drift}, our model and the MOVIN model demonstrate consistent positions at the starting and ending points when returning to the origin. In contrast, Xsens exhibits significant global drifting, confirming its final position is notably distant from the starting point.

\begin{figure}[b]
\centering
\includegraphics[width=.95\linewidth]{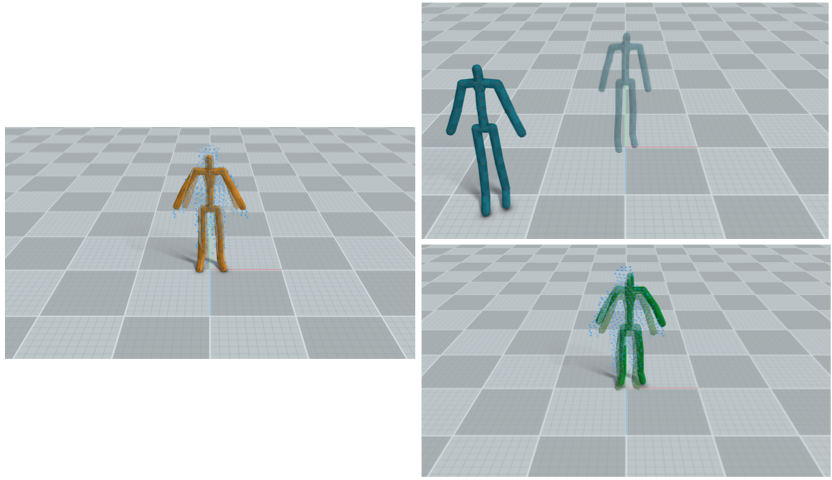}
\caption{Global Drifting results on ELMO (Yellow), Xsens (Blue), and MOVIN (Green). The initial character at the origin is depicted as transparent; upon returning to the origin, the character appears solid.}
\label{fig:drift}
\end{figure}

\subsection{Attention between joints and body-patch point groups.}
\label{subsec:attention}

In Fig.~\ref{fig:attention}, (a) shows an input point cloud and the corresponding pose output for a specific time frame. The left image of (b) visualizes the center point (highlighted in color) of 32 body-patch point groups within the input point cloud, while the right image illustrates the attention map between the right arm joints and the body-patch point groups. The x-axis of the attention map represents four joints in the right arm (grouped within the red border line in (a)), and the y-axis represents the 32 body-patch point groups.

Notably, the attention map shows high values on groups indexed 5 and 13, whose center points are located closer to the wrist and elbow joints. This suggests that our design choice of embedding the point cloud by body-patch, combined with the self-attention for Motion Tokens and Point Tokens, effectively captures the correlation between human joints and the body-patch groups from the input point cloud.

\begin{figure}[t]
\centering
\includegraphics[width=.85\linewidth]{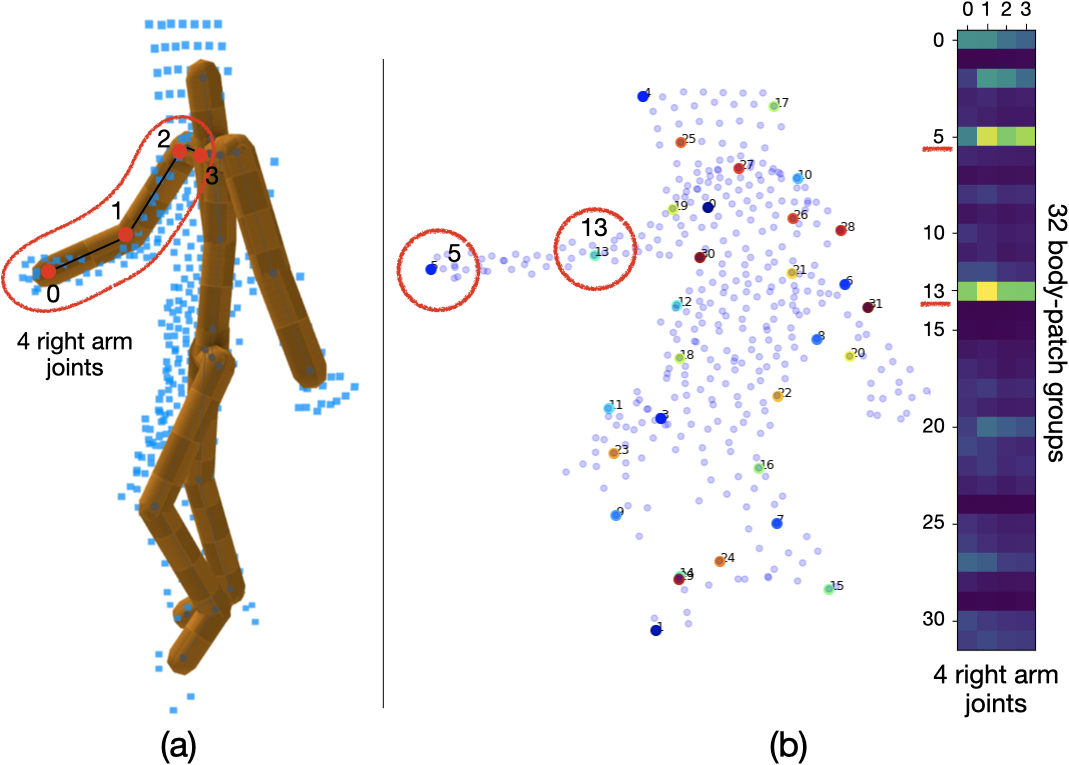}
\caption{(a) Input point cloud and output pose. (b) Attention map showing a higher correlation between right arm joints and groups indexed 5 and 13.}
\label{fig:attention}
\end{figure}

\subsection{Real-time Motion Capture Applications}
\label{subsec:live}

To highlight the practicality of ELMO, we seamlessly integrated it into real-time applications running at 60 fps using the Unity3D engine. For a comprehensive visual overview of the results, please refer to our supplementary video.

\paragraph*{\textbf{System setup.}}
The process begins by positioning a single LiDAR sensor in front of the user. A laptop processes the incoming signals from the LiDAR sensor to generate motion data output. Next, the boundary or region of interest (ROI) is defined to specify where the motion capture will occur, ensuring focus on pertinent areas. The subject's skeleton offset is then registered using our calibration model. 
Finally, real-time motion capture is initiated through the ELMO framework, enabling seamless tracking of the user's movements. Figure~\ref{fig:setup} illustrates the overall system setup process.

\begin{figure}[h]
\centering
\includegraphics[width=.95\linewidth]{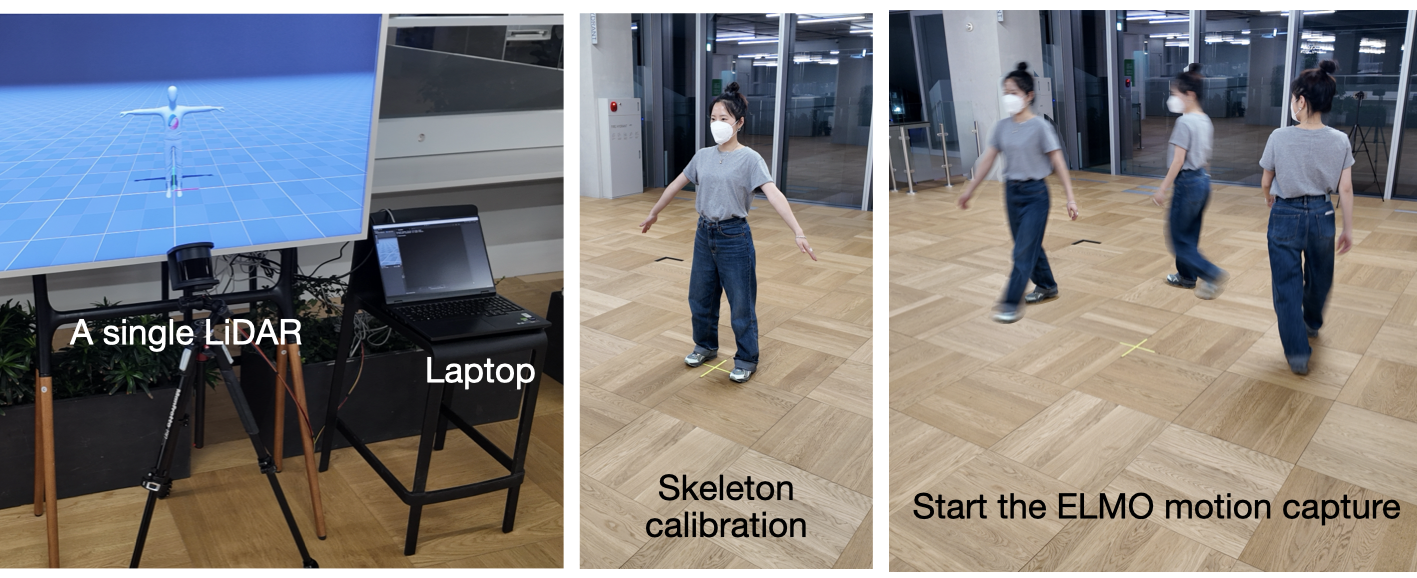}
\caption{System setup process for real-time motion capture. }
\label{fig:setup}
\end{figure}

\paragraph*{\textbf{Live streaming.}}
We demonstrate our framework's ability to stream output motion in real-time for single-subject actions, object interactions, and two-subject interactions.

Figure~\ref{fig:teaser} showcases ELMO's output poses across various actions, highlighting its versatility in capturing both common movements like walking, running, and jumping, as well as more intricate actions such as lying down, doing push-ups, and performing cartwheels. Our method accurately predicts foot contact from the output, facilitating the removal of foot-skating. Additionally, ELMO captures object interaction motions, such as sitting on a chair and placing a hand on a table. Notably, even with small occlusions, ELMO generates plausible motions, as demonstrated in the rightmost image featuring a narrow table.
Moreover, ELMO's lightweight model ensures low latency when handling multiple subjects. The left image of Fig.~\ref{fig:interaction} illustrates its proficiency in accommodating interactive dynamic actions like martial arts.

\paragraph*{\textbf{Interactive multi-player game.}} 
In the right image of Fig. \ref{fig:interaction}, two subjects are depicted engaging in a shooting game, accompanied by in-game snapshots. Our method enables precise global tracking, facilitating immersive interactions within the virtual environment.

\section{Discussion on Latency vs. Accuracy}
\begin{figure}[t]
  \centering
  \includegraphics[width=.99\linewidth]{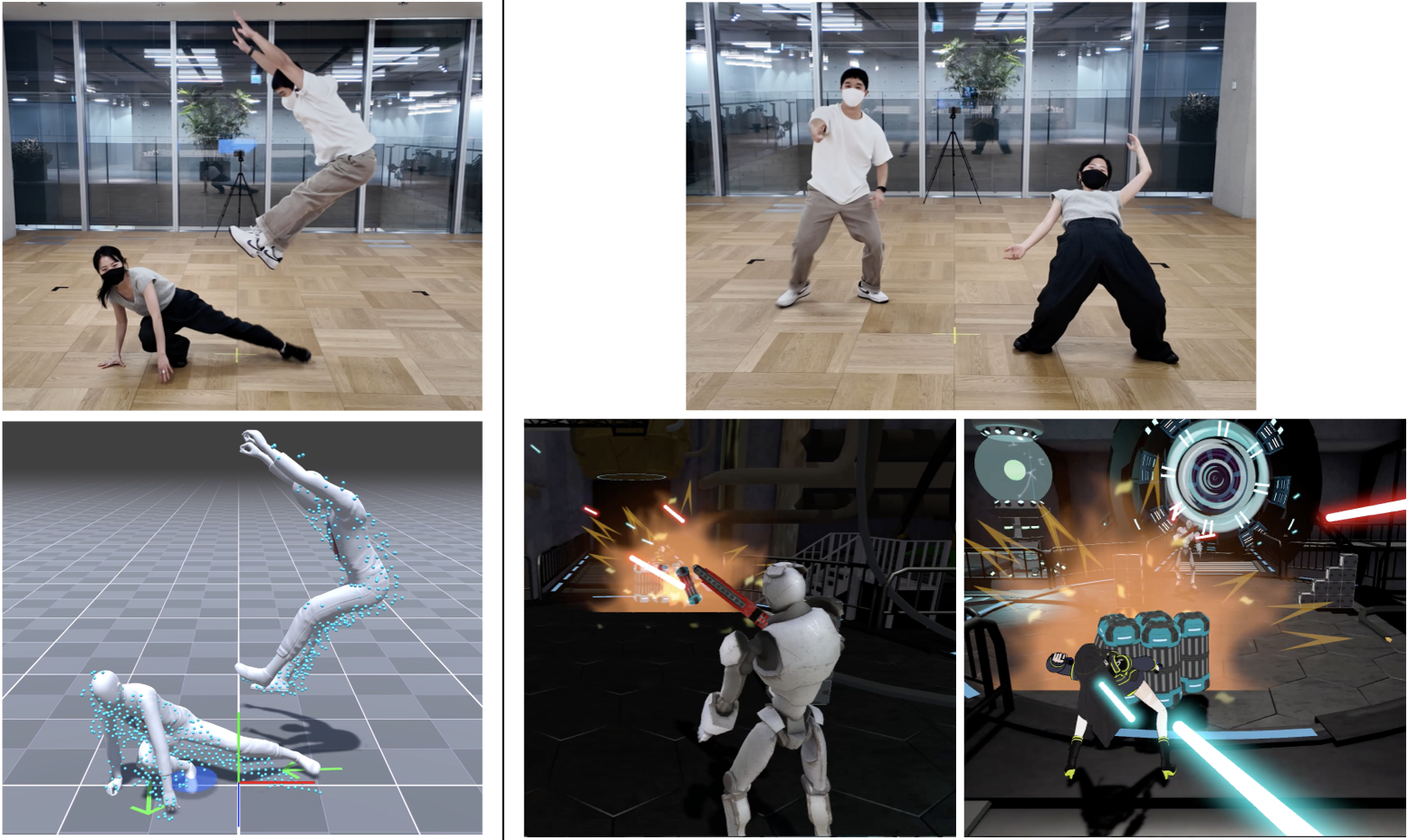}
  \caption{Left: live streaming with two subjects engaged in martial arts. Right: Interactive shooting game.}
  \label{fig:interaction}
\end{figure}

ELMO takes an in-betweening-based upsampling strategy by processing the point cloud input of frame $i+3$ and conducting a 3-frame upsampling for frames from $i+1$ to $i+3$. In contrast, the baseline ELMO (Sec.~\ref{subsec:quantitative}) predicts the poses of future frames $i+1$, $i+2$, and $i+3$ from the point cloud input of the current frame $i$, making it a prediction-based upsampling method.

While Baseline ELMO provides the benefit of zero-latency upsampling, it faces challenges when handling dynamic movements. Inaccuracies in predicting future frames can diminish accuracy, especially during acyclic rapid body motions. This can lead to accumulated errors and discontinuity in the output motions.

By utilizing the point cloud input of one future frame, ELMO introduces a latency of $33ms$ (equivalent to 2 frames at $60Hz$). However, employing in-betweening-based upsampling effectively mitigates the challenges faced by baseline ELMO and significantly improves accuracy, which is evident in our evaluation results in Section \ref{sec:Evaluation}. Given the high quality of motion capture it provides, a latency of $50ms$ is considered acceptable in real-time scenarios. Hence, we have opted for the in-betweening-based upsampling approach.

\section{Limitation and Future Work}
\label{sec:discussion}

ELMO has several limitations that need to addressed in future studies. One challenge is self-occlusion, where the subject's body obstructs other body parts, particularly when standing sideways. This can result in ELMO generating plausible but imprecise poses that may not accurately align with the actual pose. Employing a multi-LiDAR system will be a promising solution by eliminating the occluded areas.

Additionally, when the user's action significantly deviates from the distribution of the training dataset, our motion prior generates only the closest plausible motion. As a result, it may not precisely track the exact pose while producing a convincing motion. Similarly, in challenging environments with degraded input (such as missing frames/points, sparse sampling, or self-occlusions), the model is expected to generate a plausible motion but may struggle to precisely track the actual pose.

To capture multi-person scenarios, we use a clustering algorithm to separate the input point clouds from each person and perform individual motion inference. Consequently, ELMO is restricted to non-overlapping dedicated zones for each subject, as the model fails when subjects' point clouds aggregate into a single cluster.
We also handle occlusion from large objects like desks by placing the object in the desired position and filtering out its point cloud during background removal. This ensures that only the user's point cloud is captured. However, this approach cannot handle moving objects.
A promising direction for future work is integrating point cloud segmentation with the motion capture framework to address these limitation.

For data augmentation, we does not reflect variations for hair and clothing. Since users in real-world scenarios may have diverse hairstyles and outfits, our current augmentation method may not be entirely appropriate. To overcome this limitation, Combining hair and clothing with the human model could yield higher quality synthetic data.

\section{Conclusion}
\label{sec:conclusion}

We introduced ELMO, an upsampling motion capture framework utilizing a single LiDAR sensor. ELMO achieves 60 fps motion capture from a 20 fps point cloud sequence with minimal latency ($<44ms$) through its unique upsampling motion generator. Enhanced by a novel embedding module and attention coupling, ELMO delivers real-time capture performance comparable to off-the-shelf motion capture systems, eliminating the necessity for time-consuming calibration and wearable sensors. Additionally, our introduced LiDAR-Mocap data augmentation technique significantly enhances global tracking performance.

\begin{acks}
We sincerely thank the MOVIN Inc. team for their insightful discussions and invaluable assistance in capturing and processing the dataset. Sung-Hee Lee was partially supported by IITP, MSIT, Korea (2022-0-00566) and STEAM project, NRF, MIST (RS-2024-00454458).
\end{acks}

\bibliographystyle{ACM-Reference-Format}
\bibliography{reference}

\pagebreak
\begin{figure*}[t]
\centering
\includegraphics[width=0.99\textwidth]{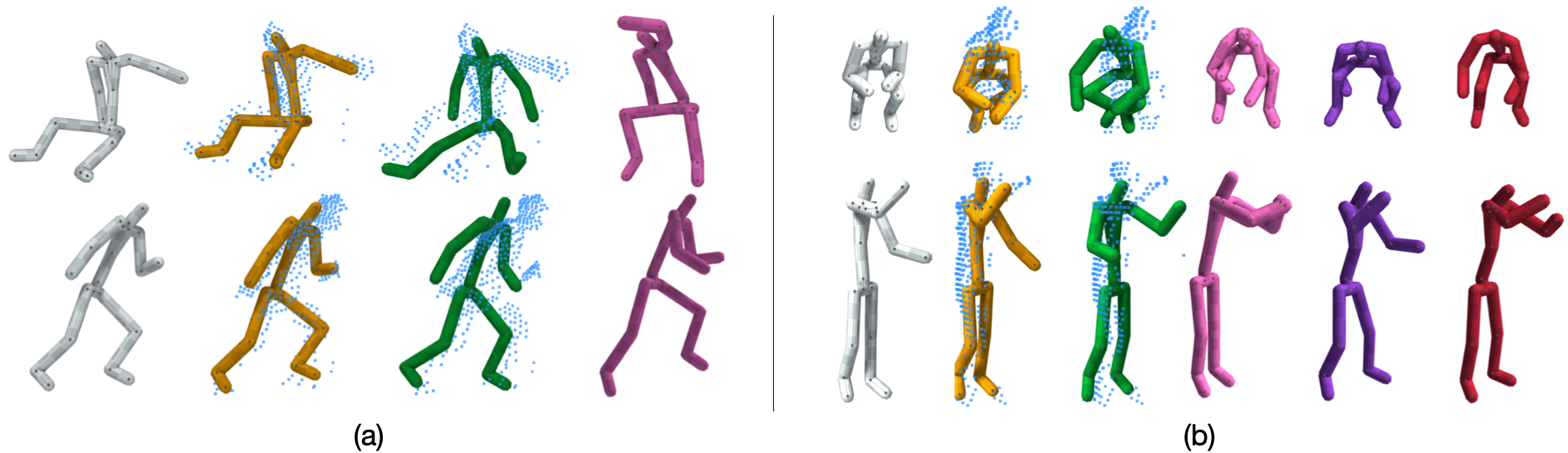}
\caption{\textbf{(a)} Samples of offline outputs on the ELMO dataset. From left to right: Ground Truth (Grey), ELMO (Yellow), MOVIN (Green), and NIKI (Pink). \textbf{(b)} Samples of offline outputs on the MOVIN dataset. From left to right: Ground Truth (Grey), ELMO (Yellow), MOVIN (Green), NIKI (Pink), MotionBert (Purple), and VIBE (Brown).}
\label{fig:quancombined}
\end{figure*}

\begin{figure*}[t]
  \centering
  \includegraphics[width=0.99\textwidth]{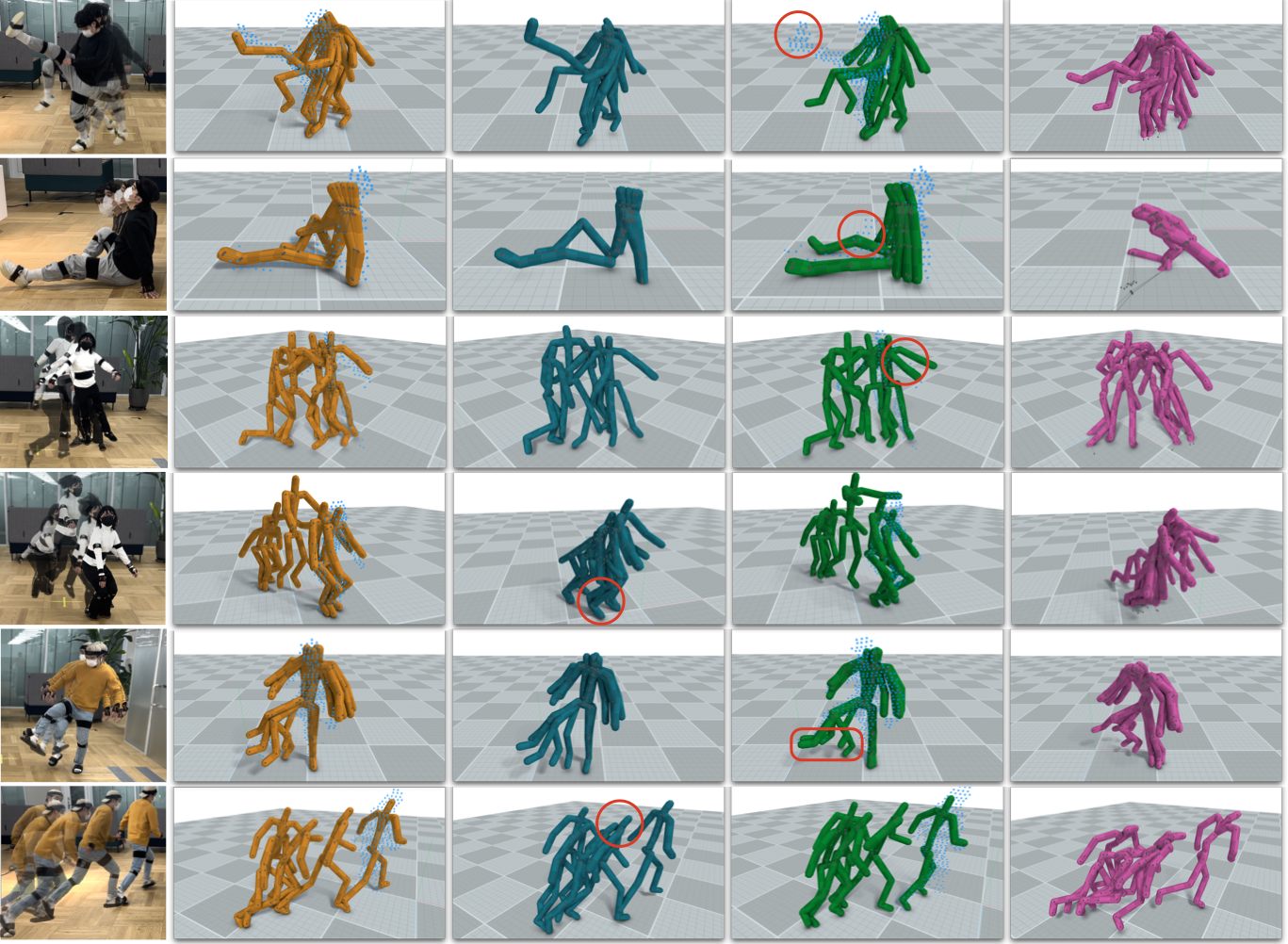}
  \caption{Qualitative comparison of real-time motion capture performance, featuring ELMO and state-of-the-art methods. The sequence, from left to right, includes ground truth, ELMO (Yellow), Xsens Awinda (Blue), MOVIN (Green), and NIKI (Pink). Notable points are denoted by red circles.}
  \label{fig:qualitative}
\end{figure*}











\end{document}